# Shock jump relations for a dusty gas atmosphere

R. K. Anand
Department of Physics, University of Allahabad, Allahabad-211002, India
E-mail: anand.rajkumar@rediffmail.com

**Abstract** This paper presents generalized forms of jump relations for one dimensional shock waves propagating in a dusty gas. The dusty gas is assumed to be a mixture of a perfect gas and spherically small solid particles, in which solid particle are continuously distributed. The generalized jump relations reduce to the Rankine-Hugoniot conditions for shocks in an idea gas when the mass fraction (concentration) of solid particles in the mixture becomes zero. The jump relations for pressure, density, temperature, particle velocity, and change-in-entropy across the shock front are derived in terms of upstream Mach number. Finally, the useful forms of the shock jump relations for weak and strong shocks, respectively, are obtained in terms of the initial volume fraction of the solid particles. The computations have been performed for various values of mass concentration of the solid particles and for the ratio of density of solid particles to the constant initial density of gas. Tables and graphs of numerical results are presented and discussed.

**Key words** Shock waves . Dusty gas . Solid particles . Shock jump relations . Mach number

## 1. Introduction

Understanding the influence of solid particles on the propagation phenomena of shock waves and on the resulting flow field is of importance for solving many engineering problems in the field of astrophysics and space science research. When a shock wave is propagated through a gas which contains an appreciable amount of dust, the pressure, the temperature and the entropy change across the shock, and the other features of the flow differ greatly from those which arise when the shock passes through a dust-free gas. The flow field, that develops when a moving shock wave hits a two-phase medium of gas and particles, has a close practical relation to industrial applications (e.g. solid rocket engine in which aluminum particles are used to reduce the vibration due to instability) as well as industrial accidents such as explosions in coalmines and grain elevators. Therefore, a successful prediction of the behavior of shock waves in a two-phase medium of gas and solid particles is very crucial and imperative for the successful design and operation of rocket nozzles and energy conversion systems. Underwater shock-wave focusing techniques such as lithotripsy for treatment of kidney stones have been widely used in hospitals. Presently, these techniques have been applied not only to urology but also musculoskeletal disorders, brain-neuro surgery, cancer treatments and the therapy of cerebral embolisms. This paper describes an interaction phenomenon when a shock wave propagates in a two-phase medium of a perfect gas and spherically small solid particles.

The objective of present study is to obtain jump relations across a shock front in a two-phase mixture of perfect gas and small solid particles. To author's best knowledge, so far there is no paper reporting the shock jump relations for a two-phase mixture of a perfect gas and small solid particles, obtained using the Pai model (1977). This model assumes the dusty gas to be a mixture of spherically small solid particles and a perfect gas. The dust phase comprises the total amount of solid particles which are continuously distributed in the perfect gas. On the one hand, the volumetric fraction of the dust lowers the compressibility of the mixture. On the other hand, the mass of the dust load may increase the total mass, and hence it may add to the inertia of the mixture. Both effects due to the addition of dust, the decrease of the mixture's compressibility and the increase of the mixture's inertia may markedly influence the shock wave spread.

In the present research paper, the author has derived the generalized forms of jump relations for one-dimensional shock wave propagating in a dusty gas. These jump relations reduce to the well known Rankine-Hugoniot conditions for shock waves in an ideal gas. The dusty gas model given by Pai (1977) is used here. The jump relations for the pressure, temperature, density, and particle velocity are obtained, respectively in terms of the upstream Mach number $M$, characterizing the shock strength. Besides these jump relations the generalized expression for the adiabatic compressibility of the mixture and change-in-entropy across the shock front are also obtained in terms of the upstream Mach number $M$. Further, the useful forms of the shock jump relations for pressure, density and mixture velocity and adiabatic compressibility of mixture along with the expression for the change-in-entropy, in terms of the initial volume fraction $Z_o$ of the solid particles are obtained for the two cases: viz., (i) when the shock is weak and (ii) when it is strong, simultaneously. The $Z_o$ is a function of $k_p$, the mass concentration of the solid particles and $G$, the ratio of the density of solid particles to the constant initial density of gas. If the parameter $Z_o$ is taken zero, the generalized shock jump relations reduce to the corresponding jump relations for shock waves in an ideal gas. These generalized jump relations for various flow variables are very useful in the theoretical and experimental investigations of strong as well as weak shock waves in the dusty environments. The background information is provided in Sect.1 as an introduction. Sect. 2 contains the general assumptions and notations. In Section 3 the generalized forms of shock jump conditions are presented. Sect. 4 mainly describes results with a brief discussion. The shock conditions corresponding to an ideal gas are summarized in an appendix for the convenience of reference.

## 2. Basic equations

The conservation equations for an unsteady, plane, cylindrically or spherically symmetric flow field between a shock and a piston moving behind it in a dusty gas under an equilibrium condition can be expressed conveniently in Eularian coordinates (*see* Anand 2012b) as follows:

$$\frac{\partial u}{\partial t} + u\frac{\partial u}{\partial r} + \frac{1}{\rho}\frac{\partial p}{\partial r} = 0 \tag{1}$$

$$\frac{\partial \rho}{\partial t} + u\frac{\partial \rho}{\partial r} + \rho\left(\frac{\partial u}{\partial r} + j\frac{u}{r}\right) = 0 \tag{2}$$

$$\frac{\partial e}{\partial t} + u\frac{\partial e}{\partial r} - \frac{p}{\rho^2}\left(\frac{\partial \rho}{\partial t} + u\frac{\partial \rho}{\partial r}\right) = 0 \tag{3}$$

where $u(r,t)$ is the velocity of the mixture, $\rho(r,t)$ the density of the mixture, $p(r,t)$ the pressure of the mixture, $e(r,t)$ the internal energy of the mixture per unit mass, $r$ is the distance from the origin, O and $t$ is the time co-ordinate. The geometry factor $j$ is defined by $j = d\,In\,A/d\,In\,r$, where $A(r) = 2\pi(j-1)r^{j-1}$ is the flow cross-section area. Then the one-dimensional flow in plane, cylindrical and spherical symmetry is characterized by $j$ = 0, 1, and 2, respectively.

Due to the condition of velocity and temperature equilibrium, the terms of drag force and heat-transfer rate, which can be expressed via the drag coefficient and the Nusselt number, do not appear in the right-hand sides of the equations (2) and (3). These terms are, of course, important for evaluating the extent of the relaxation zone behind the shock front, which is however, beyond the scope of this paper. The dusty gas is a pure perfect gas which is contaminated by small solid particles and not as a mixture of two perfect gases. The solid particles are continuously distributed in the perfect gas and in their totality are referred to as dust. It is assumed that the dust particles are highly dispersed in the gas phase such that the dusty gas can be considered as a continuous medium where the conservation Eqs. (1) - (3) apply. All relaxation processes are excluded such that no relative motion and no temperature differences between perfect gas and particles occur. The solid particles are also assumed to have no thermal motion, and, hence they do not contribute to the pressure of the mixture. As a result, the pressure $p$ and the temperature $T$ of the entire mixture satisfy the thermal equation of state of the perfect gas partition. The equation of state of the mixture subject to the equilibrium condition is given as

$$p = \frac{(1-k_p)}{1-Z}\rho R_i T \tag{4}$$

where $k_p = m_{sp}/m$, is the mass concentration of solid particles ($m_{sp}$) in the mixture ($m$) taken as a constant in the whole flow field, $Z$ is the volumetric fraction of solid particles in the mixture, $R_i$ is the gas constant and, $T$ is the temperature. The relation between $Z$ and the mass concentration of the solid particles in the mixture taken as a constant in the whole flow field is given by Pai *et al* (1980) as follows:

$$k_p = \frac{Z\rho_{sp}}{\rho} \tag{5}$$

where $Z = Z_o \rho/\rho_o$, while $\rho_{sp}$ is the species density of the solid particles and a subscript "$o$" refers to the initial values of $Z$ and $\rho$. The initial volume fraction of the solid particles $Z_o$ is, in general, not constant. But the volume occupied by the solid particles is very small because the density of the solid particles is much larger than that of the gas (*see* Miura and Glass 1985), hence, $Z_o$ may be assumed as a small constant. The initial volume fraction of small solid particles is (*see* Pai 1977)

$$Z_o = \frac{V_{sp}}{V_{go} + V_{sp}} = \frac{k_p}{G(1-k_p) + k_p}, \text{ and } \frac{Z}{\rho} = \frac{Z_o}{\rho_o} \tag{6}$$

where the volume of the mixture $V$ is the sum of the volume of the perfect gas at the reference state $V_{go}$ and the volume of the particles $V_{sp}$ which remains constant. The volumetric parameter $G$ is defined as $G = \rho_{sp}/\rho_{go}$ which is equal to the ratio of the density of the solid particles to the initial density of the gas. Hence, the fundamental parameters of the Pai model are $k_p$ and $G$ which describe the effects of the dust loading. For the dust-loading parameter $G$, we have a range of $G = 1$ to $G \to \infty$, i.e., $V_{sp} \to 0$. The variation of initial volume fraction $Z_o$ of small solid particles with mass concentration $k_p$ of solid particles in the mixture for various values of $G$ is shown in Fig.1. It is worth mentioning that $Z_o$ increases approximately linearly with increasing $k_p$ for $G \leq 10$, while it increases very slowly for values of $G \geq 50$. Fig. 2 shows the variation of the volumetric fraction $Z$ of solid particles in the mixture with $\rho/\rho_o$ for different values of $k_p$ and $G$. Note that $Z$ is proportional to the ratio of $\rho/\rho_o$, and it increases linearly with increasing $k_p$ for $G = 1$, while it increases very slowly for the values of $G \geq 10$.

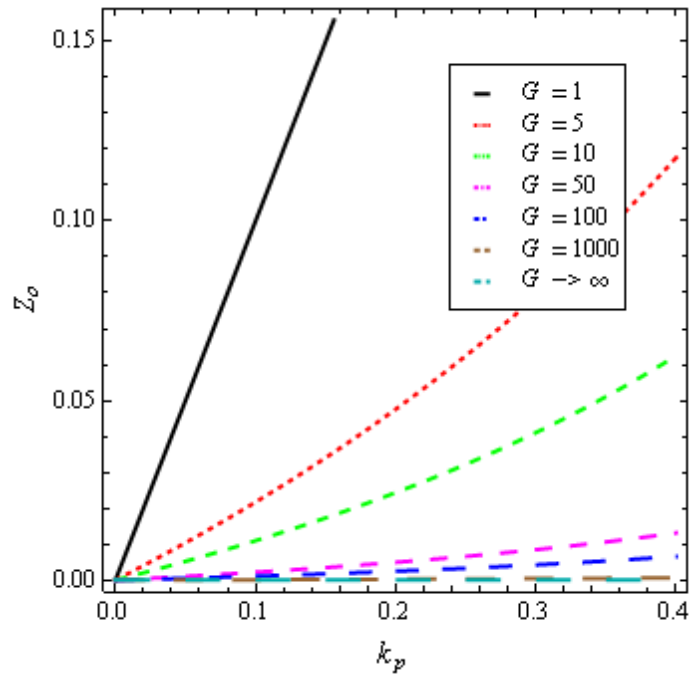

**Fig.1** Variation of $Z_o$ with $k_p$ for various values of $G$

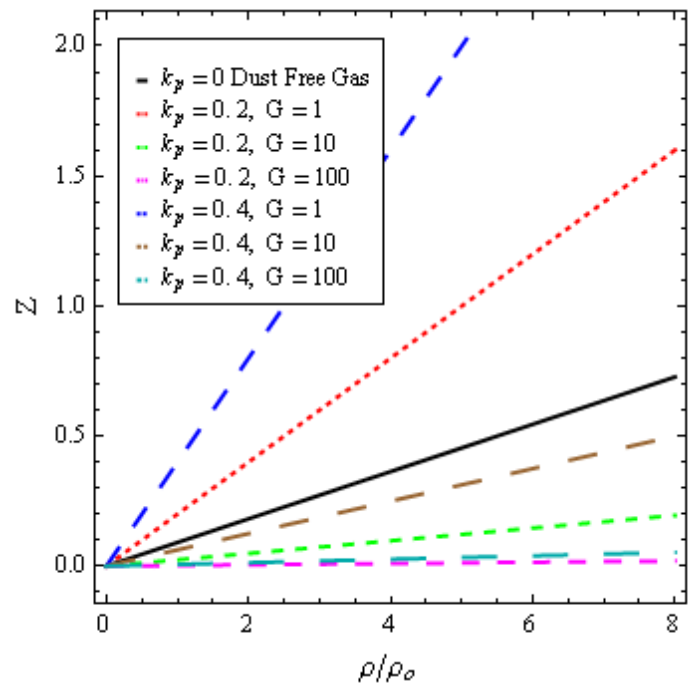

**Fig. 2** Variation of $Z$ with $\rho/\rho_o$ for various values of $k_p$ and $G$

The internal energy of the mixture is related to the internal energies of the two species and may be written as

$$e = [k_p C_{sp} + (1-k_p)C_v]T = C_{vm}T \tag{7}$$

where $C_{sp}$ is the specific heat of the solid particles, $C_v$ is the specific heat of the gas at constant volume and $C_{vm}$ is the specific heat of the mixture at constant volume. For equilibrium conditions, the specific heat of the mixture at constant pressure is

$$C_{pm} = k_p C_{sp} + (1-k_p)C_p \tag{8}$$

where $C_p$ is the specific heat of the gas at constant pressure. The ratio of the specific heats of the mixture is then given as

$$\Gamma = \frac{C_{pm}}{C_{vm}} = \frac{\gamma + \delta \beta}{1 + \delta \beta} \tag{9}$$

where $\gamma = c_p/c_v$ is the specific heat ratio of the gas, $\beta = C_{sp}/C_v$ is the specific heat ratio of the solid particles and $\delta = k_p/(1-k_p)$. The variation of the ratio of specific heat $\Gamma$ of the mixture with the mass concentration $k_p$ of solid particles in the mixture for $\beta = 1$, $\gamma = 7/5$ and various values of $\gamma$ is shown in Fig.3. It is notable that $\Gamma$ decreases with increasing $k_p$ however it increases with increasing values of $\gamma$.

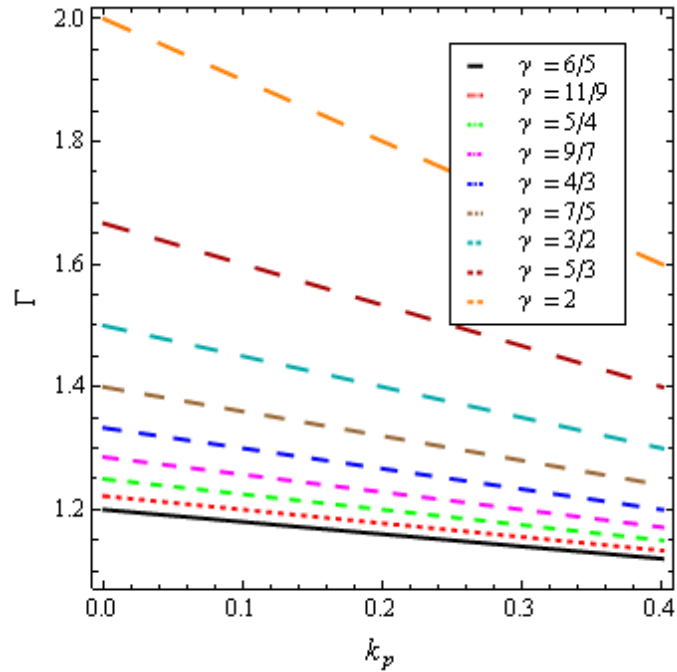

**Fig.3** Variation of $\Gamma$ with $k_p$ for $\beta = 1$ and various values of $\gamma$

Eliminating the temperature from equations (4), (6) and (7), we may now write the specific internal energy of the mixture as follows:

$$e = \left(\frac{1-Z}{\Gamma-1}\right)\frac{p}{\rho} \qquad (10)$$

For isentropic change of state of the gas-solid particle mixture and thermodynamic equilibrium condition, we can calculate the so-called equilibrium speed of sound of the mixture for a given $k_p$ by using the effective ratio of specific heats and effective gas constant $R_M = (1-k_p)R_i$ as follows:

$$a = \left(\frac{dp}{d\rho}\right)_S^{1/2} = \left(\frac{\Gamma}{(1-Z)}\frac{p}{\rho}\right)^{1/2} = \left[\frac{\Gamma(1-k_p)R_i T}{(1-Z)^2}\right]^{1/2} \qquad (11)$$

where subscript '$s$' refers to the process of constant entropy. The initial sound speed $a_o$ of the mixture is defined as

$$a_o^2 = \frac{\Gamma p_o}{(1-Z_o)\rho_o} \qquad (12)$$

Fig.4 shows the variation of non-dimensional initial sound speed $a_o^2(\rho_o/p_o)$ of the mixture with $k_p$ for $\beta=1, \gamma=7/5$ and various values of $G$. It is noteworthy that the initial sound speed increases with increasing $k_p$ for $G=1$, while it decreases approximately linearly for values of $G \geq 50$.

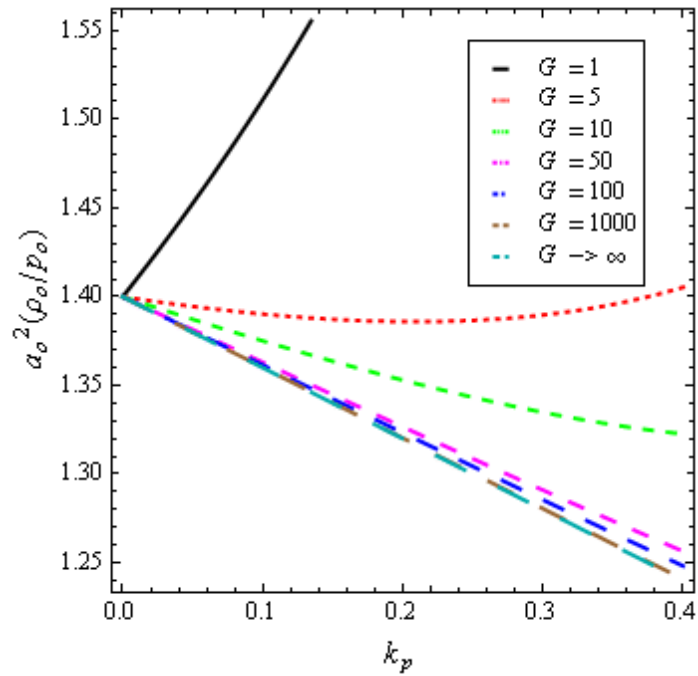

**Fig.4** Variation of $a_o^2(\rho_o/p_o)$ with $k_p$ for $\beta=1$, $\gamma=7/5$ and various values of $G$

Thus, the ratio of the equilibrium sound speed of the mixture to that of a perfect gas $a_i = (\gamma p/\rho)^{1/2}$ is

$$\frac{a}{a_i} = \frac{1}{1-Z}\left[\frac{\Gamma}{\gamma}(1-k_p)\right]^{1/2} = \frac{1}{1-Z}\left[\frac{(1+\delta\beta_{sp}/\gamma)(1-k_p)}{1+\delta\beta_{sp}}\right]^{1/2} \quad (13)$$

Fig. 5 shows the variation of $a/a_i$ with the ratio of density $\rho/\rho_o$ of the mixture for $\beta = 1$, $\gamma = 7/5$ and various values of $k_p$ and $G$. It is noteworthy that the ratio of $a/a_i$ increases exponentially with increasing $k_p$ for $G = 1$, while it decreases approximately linearly for values of $G \geq 10$.

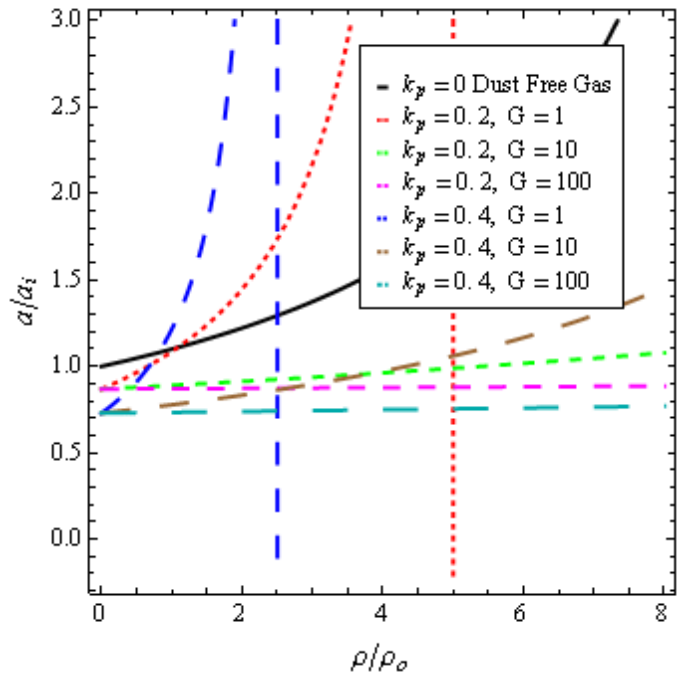

**Fig. 5** Variation of $a/a_i$ with $\rho/\rho_o$ for $\beta = 1$, $\gamma = 1.4$ and various values of $k_p$ and $G$

The deviation of the behavior of a dusty gas from that of a perfect gas is indicated by the compressibility defined as (*see* Moelwyn-Hughes 1961)

$$\tau = \frac{1}{\rho}\left(\frac{\partial\rho}{\partial p}\right)_S = \frac{(1-Z)}{\Gamma p} \quad (14)$$

where $(\partial\rho/\partial p)_S$ denotes the derivative of $\rho$ with respect to $p$ at the constant entropy. The volume of the particles lowers the compressibility of the mixture, while the mass of the solid particles increases the total mass, and therefore may add to the inertia of the mixture. This can be shown in two limiting cases of the mixture at the initial state. For $G = 1$, it follows from the equations (6) and (5) that $Z_o = k_p$, $\rho_o = p_o/R_i T$ and $\tau = (1-k_p)/\Gamma p_o$, i.e. the presence of the solid particles linearly lowers the compressibility of the mixture in the initial state. In the other limiting case, i.e. for $G \to \infty$, the volume of the solid particles $V_{sp}$ tends to zero. According to equation (6), the volume fraction $Z_o$ is equal to zero. In this case, the compressibility $\tau = 1/\Gamma p_o$ is not affected by the dust loading. The solid particles contribute only to increasing the mass and inertia of the mixture. Further, using the appropriate relations, we can write the expression for the change in entropy across the shock front as

$$\Delta s = C_{vm} \ln(p/p_o) - C_{pm} \ln(\rho/\rho_o) + C_{pm} \ln[(1-Z)/(1-Z_o)] \qquad (15)$$

where $C_{vm} = (1-k_p)R_i/(\Gamma-1)$, and $C_{pm} = \Gamma(1-k_p)R_i/(\Gamma-1)$

## 3. Formulation of jump relations in terms of Mach number

The jump conditions across shock front relate the fluid properties behind the shock, which is referred to as the downstream region, to the fluid properties ahead of the shock, which is referred to as the upstream region. The physical nature of the flow field imposes two sets of boundary conditions, and in addition, it provides also an integral relation expressing the principle of global conservation of energy inside the field. Now, let us consider a shock wave propagating into a homogeneous mixture of a perfect gas and small spherical solid particles of constant initial density $\rho_o$. In a frame of reference moving with the shock front, the jump conditions at the shock are given by the principles of conservation of mass, momentum and energy across the shock, namely,

$$\rho(U-u) = \rho_o U \qquad (16)$$

$$p + \rho(U-u)^2 = p_o + \rho_o U^2 \qquad (17)$$

$$e + \frac{p}{\rho} + \frac{(U-u)^2}{2} = e_o + \frac{p_o}{\rho_o} + \frac{U^2}{2} \qquad (18)$$

where $U$ and $u$ are, respectively, the shock front propagation velocity and the velocity of the mixture. The quantities with the suffix '$o$' and without suffix denote the values of the quantities in upstream region i.e., immediately ahead of shock, and in downstream region i.e., immediately behind the shock, respectively. Also, the effects of viscosity and thermal conductivity are omitted and it is assumed that the dusty gas has an infinite electrical conductivity. The upstream Mach number $M_{as} \approx M$ (say), which characterizes the strength of a shock, is defined as

$$M = U/a_o \qquad (19)$$

In general, the upstream Mach number $M$ in a given problem is known and it is desired to determine the downstream Mach number $M_{bs}$. The expression for downstream Mach number $M_{bs}$ can be easily obtained in terms of upstream Mach number $M$, and written as

$$M_{bs}^2 = \frac{4(1-Z_o)^4(1-Z)^2 + 2(1-Z_o)^2(1-Z)^2(\Gamma-1)M^2}{[(\Gamma+1)^2 - (1-Z_o)^2(\Gamma-1)^2]M^2 - 2(1-Z_o)^2(\Gamma-1)} \quad (20)$$

The ratio of the different flow variables such as pressure, density, temperature, etc. across a shock wave in an ideal gas are expressed as functions of upstream Mach number $M$. They are generally referred to as the Rankine-Hugoniot jump relations or conditions. Using Eqs. (10), (12), (16)-(18), the pressure of the mixture, the temperature of the mixture, the density of the mixture and the velocity of the mixture immediately behind the shock front can be written as

$$\frac{p}{p_o} = \frac{2\Gamma U^2}{(\Gamma+1)a_o^2} - \frac{\Gamma-1}{\Gamma+1} \quad (21)$$

$$\frac{T}{T_o} = \frac{[a_o^2 + (2U^2 - a_o^2)\Gamma][2a_o^2 + (\Gamma-1)U^2 + 2(U^2 - a_o^2)z_o]^2}{(1+\Gamma)^2[2a_o^2 + (\Gamma-1)U^2]U^2} \quad (22)$$

$$\frac{\rho}{\rho_o} = \frac{(\Gamma+1)U^2}{U^2(\Gamma-1+2Z_o) + 2(1-Z_o)a_o^2} \quad (23)$$

$$\frac{u}{U} = \frac{2(1-Z_o)(U^2 - a_o^2)}{(\Gamma+1)U^2} \quad (24)$$

The pressure of the mixture, the temperature of the mixture, the density of the mixture and the velocity of the mixture just behind the shock front in terms of the upstream Mach number $M$ can be expressed as

$$\frac{p}{p_o} = \frac{2\Gamma M^2 - (\Gamma-1)}{(\Gamma+1)} \quad (25)$$

$$\frac{T}{T_o} = \frac{[1 + (2M^2 - 1)\Gamma][2 + (\Gamma-1)M^2 + 2(M^2 - 1)z_o]^2}{(1+\Gamma)^2[2 + (\Gamma-1)M^2]M^2} \quad (26)$$

$$\frac{\rho}{\rho_o} = \frac{(\Gamma+1)M^2}{M^2(\Gamma-1+2Z_o) + 2(1-Z_o)} \quad (27)$$

$$\frac{u}{a_o} = \frac{2(1-Z_o)(M^2 - 1)}{M(\Gamma+1)} \quad (28)$$

The expression for the compressibility of mixture just behind the shock front is easily obtained by substituting Eq. (25) into Eq. (14) and it can be written as

$$\tau(p_o) = \frac{(1-Z)(\Gamma+1)}{\Gamma[2\Gamma M^2 - (\Gamma-1)]} \quad (29)$$

Further using Eqs. (25) and (27) in Eq. (15) the expression for the change in entropy across the shock front can easily be written as

$$\frac{\Delta s}{R_i} = \frac{(1-k_p)}{\Gamma-1} \ln\left(\frac{2\Gamma M^2 - (\Gamma-1)}{\Gamma+1}\right) - $$

$$\frac{\Gamma(1-k_p)}{\Gamma-1} \ln\left(\frac{(\Gamma+1)M^2}{M^2(\Gamma-1+2Z_o)+2(1-Z_o)}\right) + \frac{\Gamma(1-k_p)}{\Gamma-1} \ln\left(\frac{1-Z}{1-Z_o}\right) \quad (30)$$

The above forms of the jump relations across a shock front in a dusty gas flow are similar to the well-known Rankine-Hugoniot conditions across a shock front in an ideal gas flow [*see* Appendix] and the shock jump relations are explicitly written in terms of the upstream parameters only.

**3.1 Jump relations for weak shocks:** For weak shock, $M$ is taken as $M = 1+\varepsilon$, where $\varepsilon$ is a parameter which is negligible in comparison with unity i.e., $\varepsilon \ll 1$. Thus, the pressure, the density, the velocity of the mixture, and the sound speed just behind the weak shock can be, respectively, written as

$$\frac{p}{p_o} = 1 + \frac{4\Gamma\varepsilon}{\Gamma+1} \quad (31)$$

$$\frac{\rho}{\rho_o} = 1 + \frac{4(1-Z_o)\varepsilon}{\Gamma+1} \quad (32)$$

$$\frac{u}{a_o} = \frac{4(1-Z_o)\varepsilon}{\Gamma+1} \quad (33)$$

$$a \approx a_o \quad (34)$$

**3.2 Jump relations for strong shocks:** For strong shock, $U \gg a_o$, thus the pressure, the density, the velocity of the mixture, and the sound speed just behind the strong shock can be, respectively, written as

$$p = \frac{2\rho_o(1-Z_o)U^2}{\Gamma+1} \quad (35)$$

$$\rho = \frac{\rho_o(\Gamma+1)}{(\Gamma-1+2Z_o)} \quad (36)$$

$$u = \frac{2(1-Z_o)U}{(\Gamma+1)} \quad (37)$$

$$a = \frac{\sqrt{2\Gamma(\Gamma-1+2Z_o)}\,U}{(\Gamma+1)} \quad (38)$$

**3.4 Strength of a shock wave:** In shock wave analysis, the quantity $\xi = (p-p_o)/p_o = p/p_o - 1$, represents the strength of the shock, and $\Delta p = p - p_o$ denotes the overpressure. The shock speed increases with increasing overpressure.

Substituting for $p/p_o$ from equation (23), we get

$$\xi = \frac{2\Gamma}{\Gamma+1}M^2 - \frac{\Gamma-1}{\Gamma+1} - 1$$

$$\xi = \frac{2\Gamma}{\Gamma+1}(M^2 - 1)$$

Thus, for shocks of any strength, we can write

$$\xi \propto (M^2 - 1) \quad \text{i.e.,} \quad \xi \propto (M^2 - const.)$$

**Case I. For Shocks of vanishing strength:** Shock waves for which $\xi$ is almost zero, are referred to as shocks of vanishing strength. For such shocks, $M = 1$.

Thus for shocks of vanishing strength we can write, $p/p_o \approx 1$, $\rho/\rho_o \approx 1$, $T/T_o \approx 1$, and $\Delta s/R_i \approx 0$.

**Case II. For Strong shocks:** Since shock strength is proportional to $(M^2 - const.)$, strong shocks are a result of very high values of upstream Mach number $M$. The maximum values of quantities in equations (22)-(24), are given below

$$\lim_{M \to \infty} p/p_o = \infty$$

$$\lim_{M \to \infty} \rho/\rho_o = (\Gamma+1)/(\Gamma-1+2z_o)$$

$$\lim_{M \to \infty} T/T_o = \infty$$

## 4 Results and discussion

The simplified forms of the shock jump relations for one-dimensional shock waves propagating in a dusty gas are derived which reduce to the well known Rankine-Hugoniot shock conditions for ideal gas when mass concentration of solid particles in the mixture becomes zero. The jump relations for pressure, temperature, density, and velocity of mixture are obtained, respectively in terms of the upstream Mach number. The expressions for the compressibility of the mixture which shows the deviation of the behavior of a dusty gas from that of a perfect gas and the change in entropy across the shock front are also obtained, respectively in terms of the upstream Mach number. Further, the useful forms of shock jump relations for pressure, density and particle velocity in terms of the initial volume fraction $Z_o$ of small solid particles and the ratio of specific heats of the mixture $\Gamma$ are obtained for the two cases: viz., (i) when the shock is weak and (ii) when it is strong, simultaneously. For the purpose of numerical calculations, the values of the $\beta$ and the ratio of specific heats of the gas $\gamma$ are taken to be 1 and 1.4, respectively. The values $\beta = 1$ and $\gamma = 1.4$ corresponds to the mixture of air and glass particles (Miura and Glass, 1985). In our analysis, we have assumed the initial volume fraction $Z_o$ of solid particles to be a small constant. The values of $k_p = 0.2, 0.4$ with $G = 1, 10, 100$ give small values of $Z_o$. The expression for the downstream Mach number $M_{bs}$ is given by equation (20). Fig. 6 depicts the variation of downstream Mach number $M_{bs}$ with upstream Mach number $M$ for $\beta = 1, \gamma = 1.4$ and various values of $k_p$ and $G$. It is notable that the downstream Mach number $M_{bs}$ decreases with increasing $M$ and $k_p$, respectively while it increases with increasing $G$. This behavior, especially for the case of $k_p = 0.4$ and $G = 1$, differs greatly from the dust-free case.

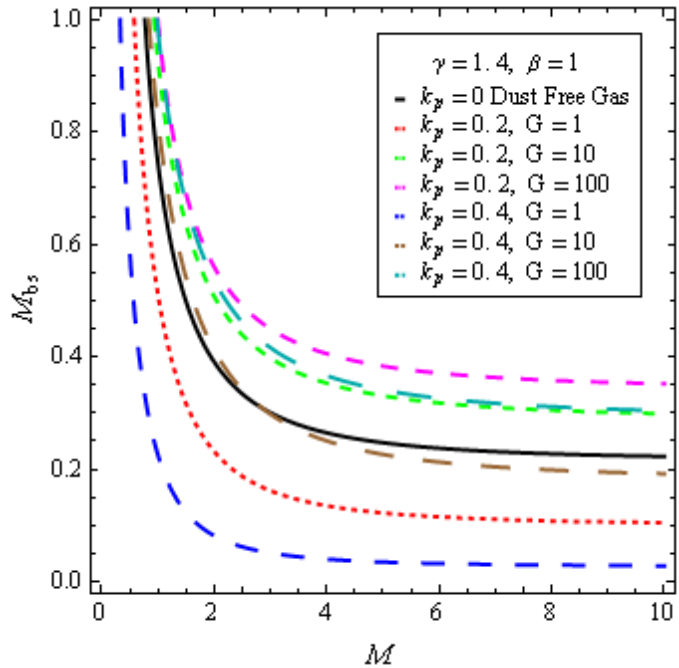

**Fig. 6** Variation of $M_{bs}$ with $M$ for various values of $k_p$ and $G$

The simplified shock jump relations for the pressure $p/p_o$, the temperature $T/T_o$, the density $\rho/\rho_o$, the velocity $u/a_o$ of mixture, the compressibility $\tau(p_o)$ of mixture just behind the shock front are given by Eqs. (25) - (29), respectively. To see the effect of the mass concentration $k_p$ and the mass-loading $G$ of the dust on the flow field behind the shock front, the plots of dimensionless pressure, temperature, density, particle velocity, sound speed and compressibility with upstream Mach number $M$ are presented in Fig. 7 for $\beta = 1, \gamma = 1.4, k_p = 0, k_p = 0.2, G = 1, k_p = 0.2, G = 10, k_p = 0.2, G = 100,$ $k_p = 0.4, G = 1, k_p = 0.4, G = 10, k_p = 0.4, G = 100$, allowing a comparison with the dust-free case for $k_p = 0$. It is important to note that the pressure, temperature, density, velocity of mixture and sound speed increases with increasing $M$ while the compressibility decreases. The pressure decreases slowly with increasing $k_p$ while it is independent of $G$. This behavior of pressure, especially for the case of $k_p = 0.4$ differs greatly from the dust-free case $k_p = 0$. The temperature increases with increasing $k_p$ for values of $G \leq 10$, while it decreases for $G = 100$ and it decreases with increasing $G$. This behavior of temperature, especially for the case of $k_p = 0.4$ and $G = 1$ differs greatly from the dust-free case. The density behaves inversely; it decreases with increasing $k_p$ for $G \leq 10$, while it increases for $G = 100$ and it increases with increasing $G$. This behavior of density, especially for the case of $k_p = 0.4$ and $G = 100$ differs greatly from the dust-free case. The velocity of mixture decreases with increasing $k_p$ while it increases with increasing $G$. This behavior of velocity, especially for the case of $k_p = 0.4$ and $G = 1$ differs greatly from the dust-free case. The speed of sound increases with increasing $k_p$ for values of $G \leq 10$, while it decreases for $G = 100$ and it decreases with increasing $G$. This behavior of sound speed, especially for the case of $k_p = 0.4$ and $G = 1$ differs greatly from the dust-free case. The compressibility decreases with increasing $k_p$ while it increases with increasing $G$. This behavior of compressibility, especially for the case of $k_p = 0.4$ and $G = 1$ differs greatly from the dust-free case. It is worth mentioning that the compressibility decreases rapidly for values of $M \leq 4$ and then it becomes constant for values of $M < 4$. It is interesting to note that the increase in pressure and temperature can be infinitely large for sufficiently large shock strengths (or Mach number) but the density increase is limited to the value $(\Gamma + 1)/(\Gamma - 1 + 2z_o)$.

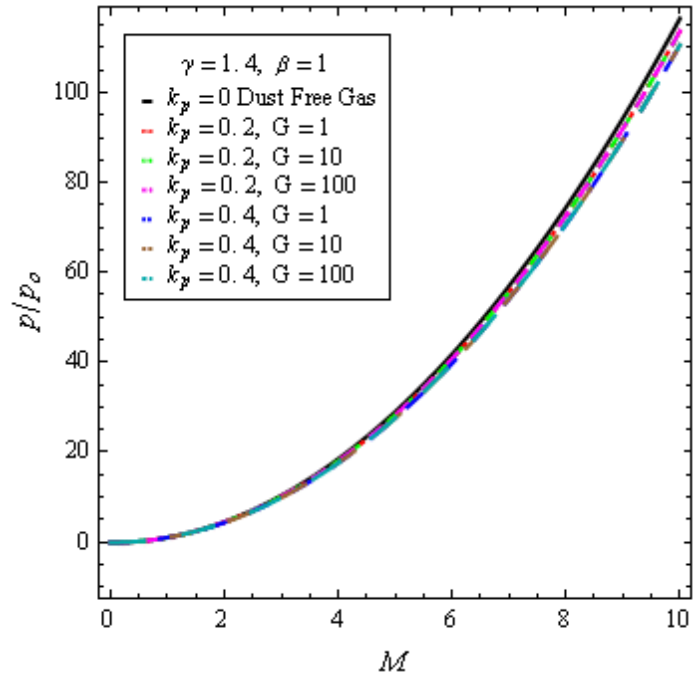

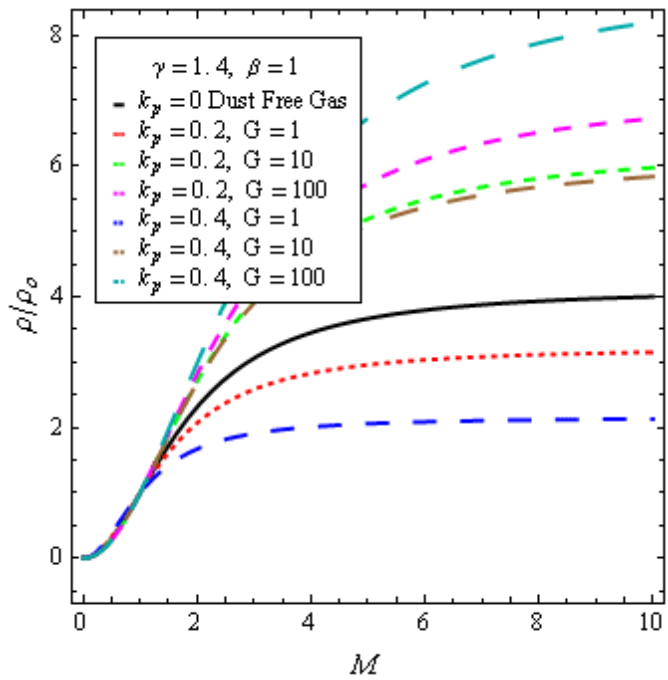

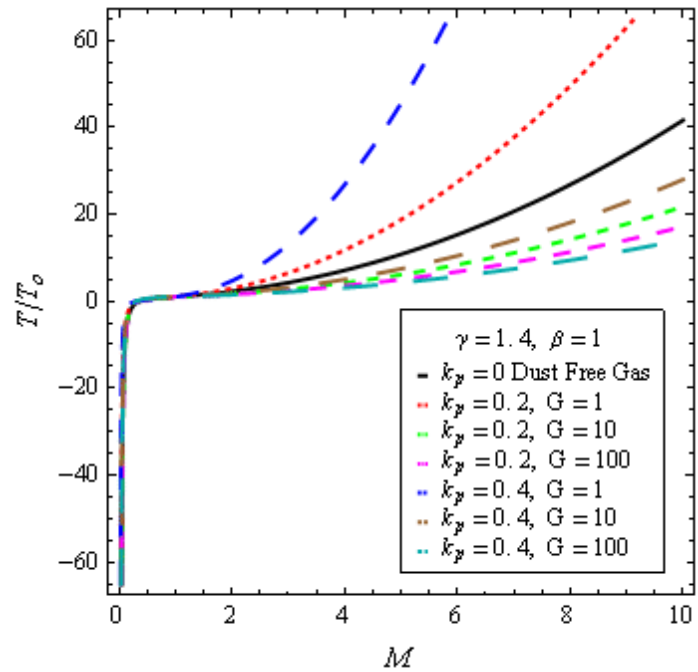

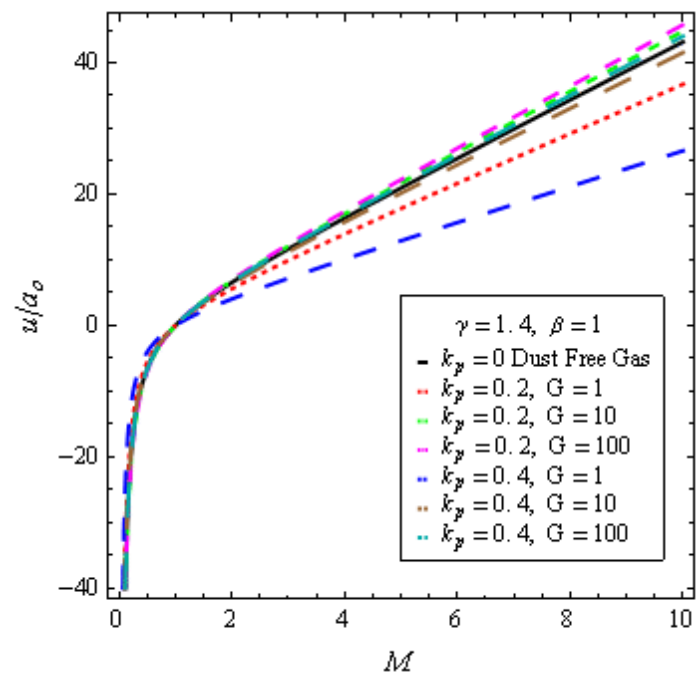

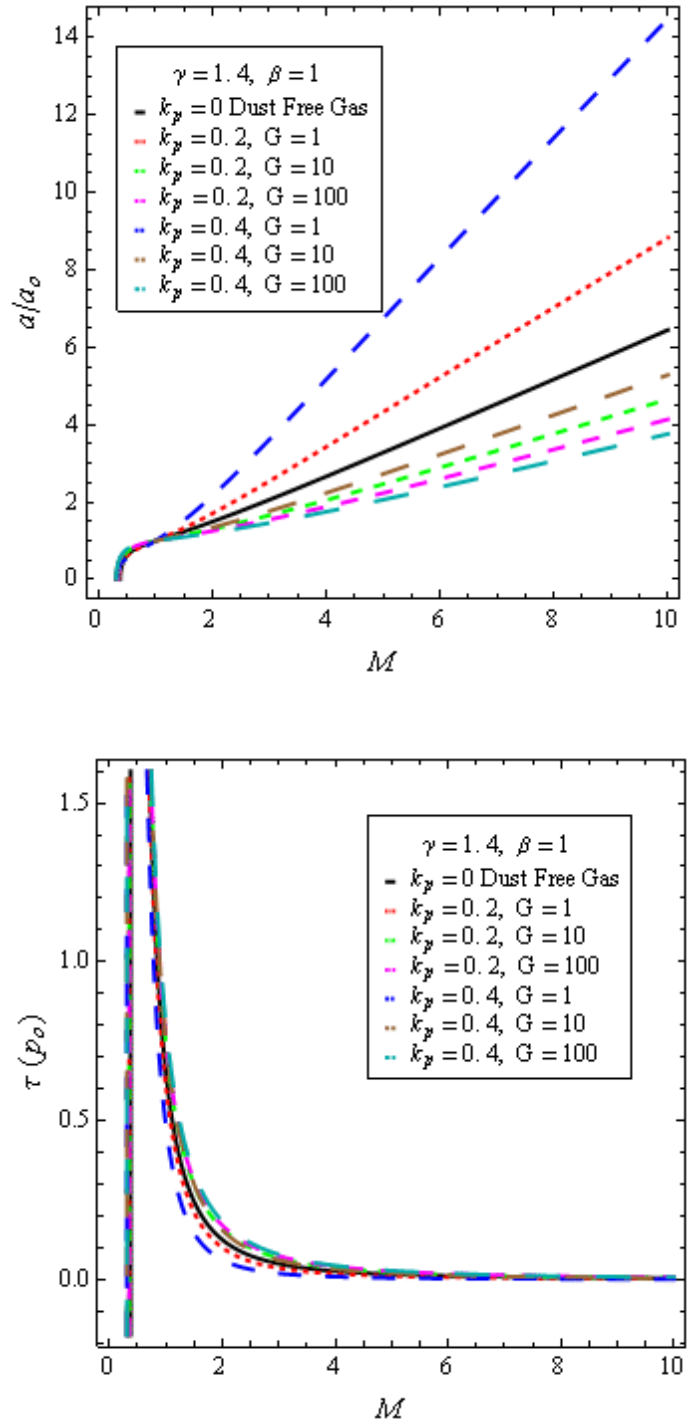

**Fig.7** Variations of $p/p_o, \rho/\rho_o, T/T_o, u/a_o, a/a_o$ and $\tau(p_o)$ with $M$ for various values of $k_p$ and $G$

The expression for the change in entropy $\Delta s/R_i$ across the shock front in a dusty gas is given by Eq. (30). The variation of the change in entropy $\Delta s/R_i$ with $M$ for $\beta = 1$, $\gamma = 1.4$ and various values of $k_p$ and $G$ are shown in Fig. 8. The change in entropy increases with increasing $M$. It is notable that with increasing $k_p$ the change in entropy increases rapidly for $G = 1$, moderately for $G = 10$ and very slightly for $G = 100$ and it decreases with increasing $G$. In other words, with increasing $k_p$ the shock becomes stronger swiftly for lower values of $G$ and gradually for higher values of $G$. According to the second law of thermodynamics, the entropy of a substance cannot be decreased by internal processes alone, thus the downstream entropy in a shock must equal or exceed its upstream value, $s \geq s_o$. This entropy increase, predicted by the mass, momentum, and energy conservation relations alone, implies an irreversible dissipation of energy, even for an ideal fluid, entirely independently of the existence of a dissipation mechanism. The expression for the change in entropy $\Delta s/R_i$ across the shock front in a dusty gas is given by Eq. (30). For the sake of justification the Eq. (30), i.e. $\Delta s/R_i$ is plotted with $M$ for $\beta = 1, \gamma = 1.4$ and various values of $k_p$ and $G$ and shown in Fig. 8. The change in entropy increases with increasing $M$. It increases with increasing $k_p$ for values of $G \leq 10$, while it remains the same for $G = 100$ and it decreases with increasing $G$. This behavior, especially for the case of $k_p = 0.4$ and $G = 1$ differs greatly from the dust-free case. Obviously, the change in entropy for ideal gas ($k_p = 0$) is positive when upstream Mach number $M$ is greater than unity; hence, only a shock from supersonic to subsonic speed is possible, with a corresponding rise in pressure across the normal discontinuity. It is important to note that for curves except for $k_p = 0$ (dust free gas) $\Delta s/R_i$ has always positive values for values of $M$ greater than unity. Decrease in entropy is impossible by second law of thermodynamics, thus shock waves cannot develop in a flow where upstream Mach number is less than one in case of a dusty gas as well as in an ideal gas. It confirms that the shock waves will arise in a flow of dust-laden fluids where upstream Mach number is equal to or greater than unity as in the flow of ideal gas. It is remarkable and astonishing that in non-ideal gas flows the shock waves may arise where Mach number equal to or greater than 0.5 (Anand, 2012a).

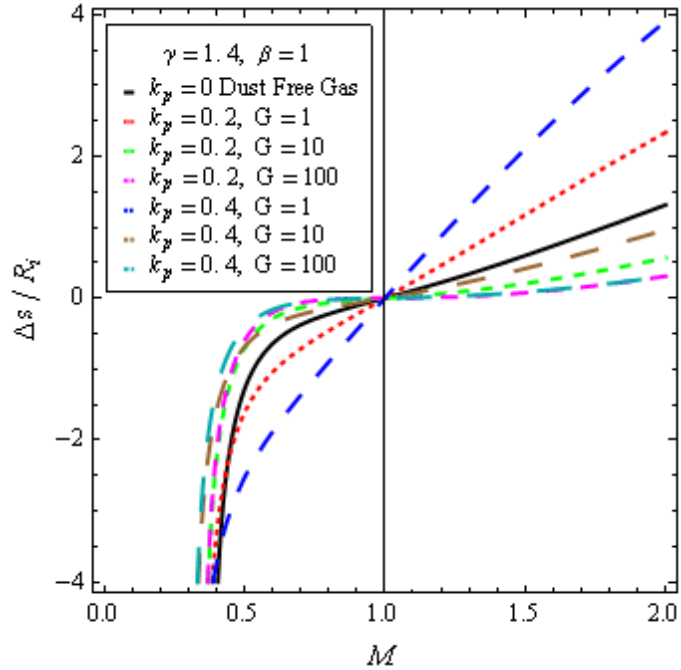

**Fig. 8** Variations of $\Delta s/R_i$ with $M$ for various values of $k_p$ and $G$

**4.1 Weak shock waves** The handy forms of jump relations for weak shock waves are given by Eqs. (31)-(34). These relations are dependent of $\varepsilon(r)$ a parameter which is negligible in comparison with the unity, the ratio of specific heat $\Gamma$ of the mixture, the mass concentration $k_p$ of solid particles and the dust laden parameter $G$. The variations of the pressure $p/p_o$, density $\rho/\rho_o$, velocity $u/a_o$ of mixture and change in entropy with $\varepsilon$ for $\beta = 1, \gamma = 1.4$ and various values of $k_p$ and $G$ are shown in Fig. 9. It is important to note that the pressure, density, velocity of the mixture and change in entropy increase with increasing $\varepsilon$. The pressure decreases very small in amount with increasing $k_p$ and it is independent of $G$. This behavior of pressure, especially for the case of $k_p = 0.4$ differs greatly from the dust-free case. The density decreases with increasing $k_p$ for values of $G \leq 10$, while it increases for $G = 100$ and it increases with increasing $G$. This behavior of density, especially for the case of $k_p = 0.4$ and $G = 1$ differs greatly from the dust-free case. The velocity of mixture also decreases with increasing $k_p$ for values of $G \leq 10$, while it increases for $G = 100$ and it increases with increasing $G$. This behavior of velocity, especially for the case of $k_p = 0.4$ and $G = 1$ differs greatly from the dust-free case. The change in entropy slightly increases with increasing $k_p$ and it also slightly decreases with increasing $G$. This behavior of change in entropy, especially for the case of $k_p = 0.4$ and $G = 1$ differs greatly from the dust-free case.

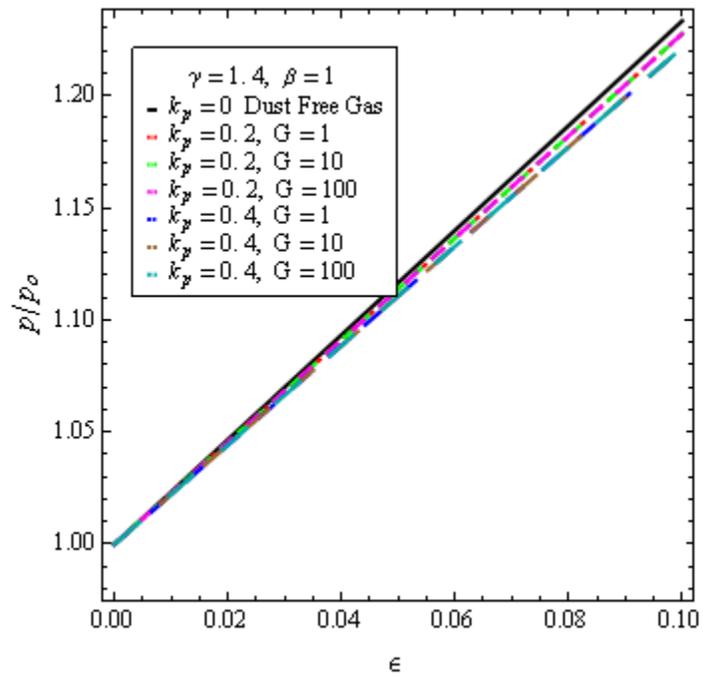

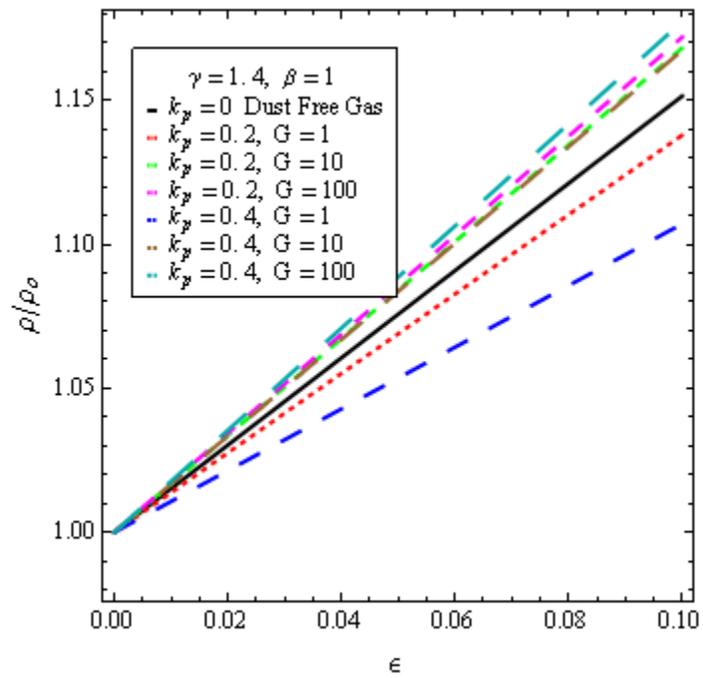

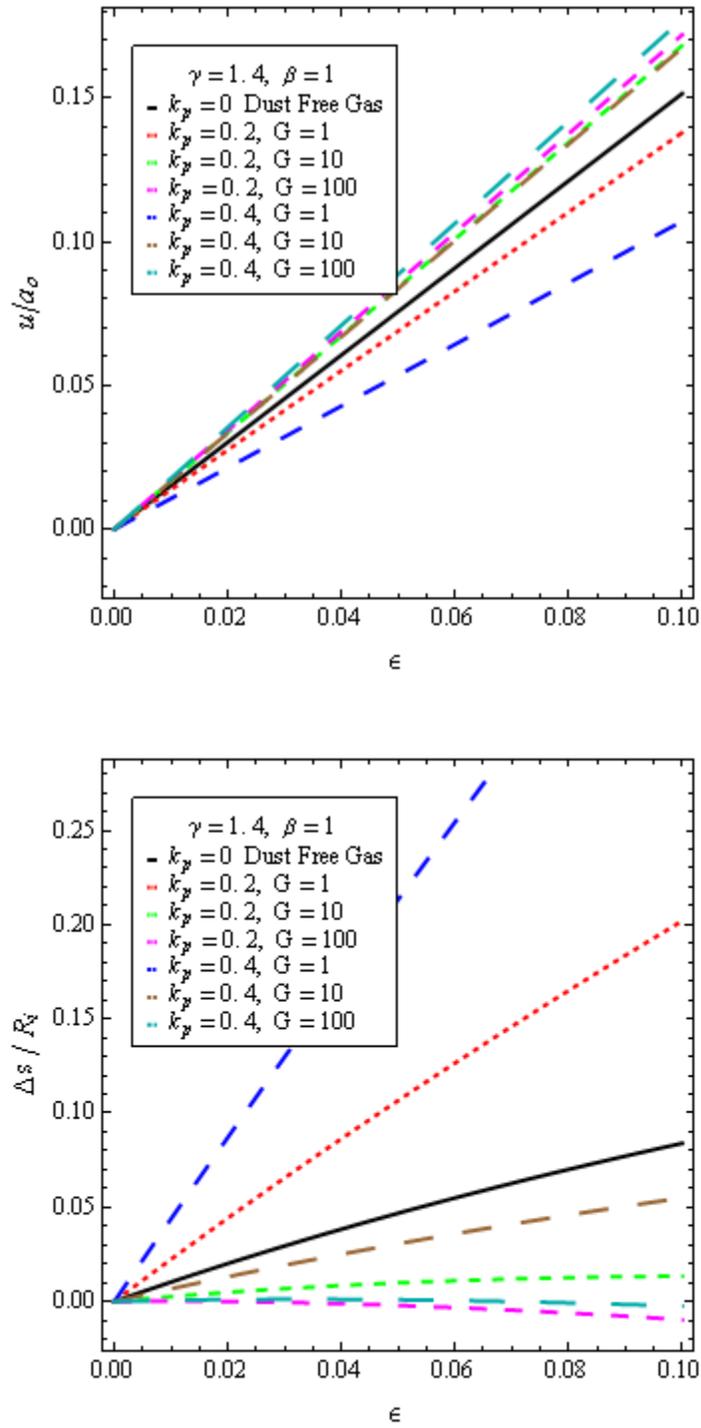

**Fig.9** Variations of $p/p_o$, $\rho/\rho_o$, $u/a_o$ and $\Delta s/R_i$ with $\varepsilon$ for various values of $k_p$ and $G$

**4.2 Strong shock waves** The handy forms of jump relations for strong shock waves are given by Eqs. (35)-(38). These jump relations are dependent of the shock strength $U/a_o$, the ratio of specific heat $\Gamma$ of the mixture, the mass concentration $k_p$ of solid particles and the dust laden parameter $G$. Fig. 10 shows the variations of the pressure $p/p_o$, density $\rho/\rho_o$, velocity $u/a_o$ of mixture, speed of sound $a/a_o$ and change in entropy with non-dimensional shock velocity for $\beta = 1, \gamma = 1.4$ and various values of $k_p$ and $G$. It is important to note that the pressure, velocity of mixture, speed of sound and change in entropy increase with increasing $U/a_o$ but the density remains unchanged. The pressure decreases very small in amount with increasing $k_p$ and it is independent of $G$. This behavior of pressure, especially for the case of $k_p = 0.4$ differs greatly from the dust-free case. The density decreases with increasing $k_p$ for values of $G \leq 10$, while it increases for $G = 100$ and it increases with increasing $G$. This behavior of density, especially for the case of $k_p = 0.4$ and $G = 100$ differs greatly from the dust-free case. The velocity of mixture remains unchanged with increasing $k_p$ for $G = 1$, while it increases slightly for values of $G \geq 10$ and it increases with increasing $G$. This behavior of velocity, especially for the case of $k_p = 0.4$ and $G = 100$ differs greatly from the dust-free case.

The speed of sound increases with increasing $k_p$ for $G = 1$, remains unchanged for $G = 10$ decreases for $G = 100$ and it increases with increasing $G$. This behavior of speed, especially for the case of $k_p = 0.4$ and $G = 100$ differs greatly from the dust-free case.

The change in entropy increases with increasing $k_p$ for values of $G \leq 10$, while it slightly decreases for $G = 100$ and it decreases with increasing $G$. This behavior of change in entropy, especially for the case of $k_p = 0.4$ and $G = 1$ differs greatly from the dust-free case.

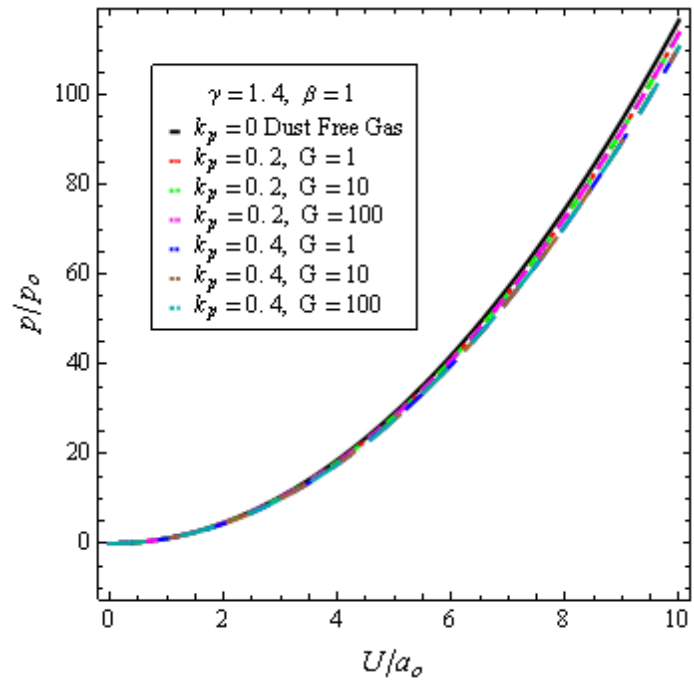

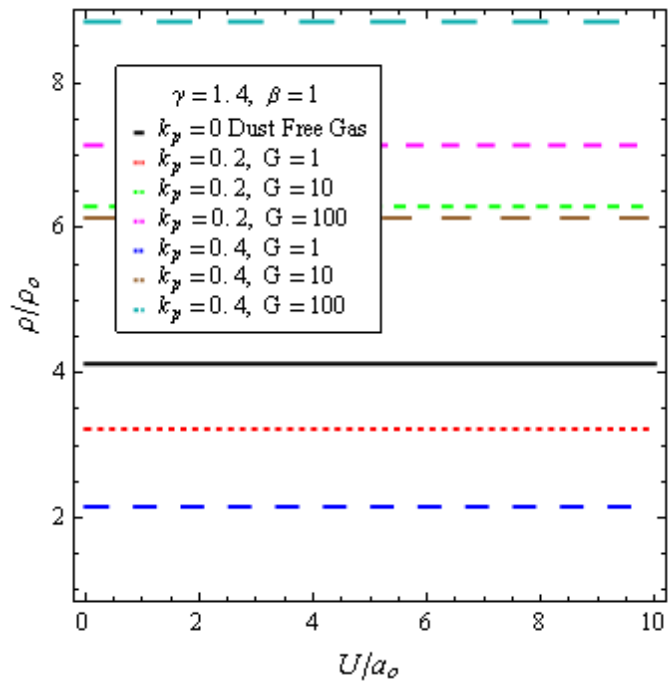

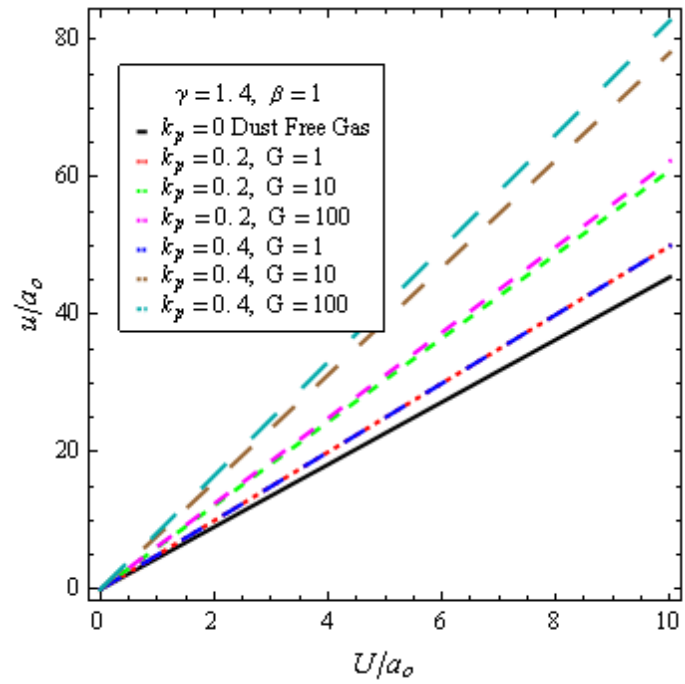

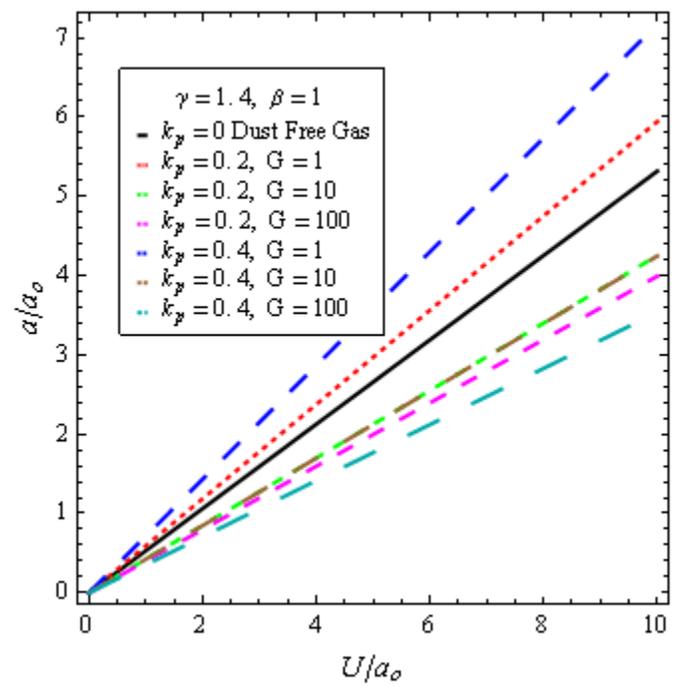

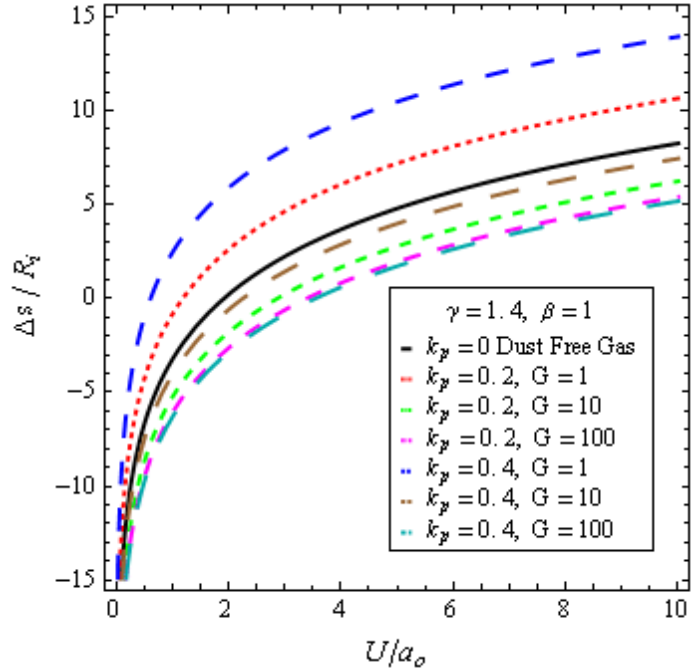

**Fig.10** Variations of $p/p_o, \rho/\rho_o, u/a_o, a/a_o$ and $\Delta s/R_i$ with $U/a_o$ for various values of $k_p$ and $G$

**Acknowledgement:** I acknowledge the encouragement of my wife, Nidhi during preparation of the paper.

**Appendix** Shock jump relations for ideal gas (Anand 2000)

$$p = \rho_o\, a_o^2 \left\{ \frac{2M^2}{\gamma+1} - \frac{\gamma-1}{\gamma(\gamma+1)} \right\}$$

$$\rho = \rho_o \frac{(\gamma+1)M^2}{(\gamma-1)M^2 + 2}$$

$$T = \frac{T_o\left[2+(\gamma-1)M^2\right]\left\{2\gamma M^2 - (\gamma-1)\right\}}{(\gamma+1)^2 M^2}$$

$$u = a_o \frac{2}{\gamma+1}\left(M - \frac{1}{M}\right)$$

$$\frac{\Delta s}{R} = \frac{\gamma}{(\gamma-1)} \ln\left[\frac{2+(\gamma-1)M^2}{(\gamma+1)M^2}\right] + \frac{1}{\gamma-1}\ln\left[\frac{2\gamma M^2}{(\gamma+1)} - \frac{(\gamma-1)}{(\gamma+1)}\right]$$

where $M = U/a_o$ and $a_o^2 = \gamma\, p_o/\rho_o$

Jump relations for weak shocks in ideal gas

$$p = p_o\left\{1+\frac{4\gamma\varepsilon}{\gamma+1}\right\},\; \rho = \rho_o\left\{1+\frac{4\varepsilon}{\gamma+1}\right\},\; T = T_o\left\{1+\frac{8\gamma(\gamma-1)\varepsilon}{(\gamma+1)^2}\right\},$$

$$u = \frac{4a_o}{\gamma+1}\varepsilon \text{ and } U = a_o(1+\varepsilon)$$

Jump relations for strong shocks in ideal gas

$$p = \frac{2}{\gamma+1}\rho_o U^2,\; \rho = \frac{(\gamma+1)}{(\gamma-1)}\rho_o,\; T = 2T_o\frac{(\gamma-1)\gamma}{(\gamma+1)^2}\left(\frac{U}{a_o}\right)^2,$$

$$u = \frac{2}{\gamma+1}U \text{ and } a = \frac{\sqrt{2\gamma(\gamma-1)}}{\gamma+1}U$$